\documentclass[11pt]{article}
\usepackage[margin=1in]{geometry}
\usepackage{amsmath}
\usepackage{amssymb}
\usepackage{booktabs}
\usepackage{array}
\usepackage{footnote}
\usepackage{arydshln}
\usepackage{float}
\usepackage{graphicx}
\usepackage{epigraph}
\usepackage[authoryear,round]{natbib}
\makesavenoteenv{tabular}
\usepackage{url}
\usepackage{setspace}
\title{Public Infrastructure Investments for Space Market Development}
\author{Akhil Rao\thanks{Rational Futures, Washington, D.C. Email: akhil@rationalfutures.com.}
\thanks{I am grateful to Alex MacDonald, Adrian Mangiuca, Moon Kim, Tom Colvin, Renata Kommel, Elaine Gresham, Carie Mullins, and Cullen Balinski for many helpful discussions and insights. All errors are my own.}}
\date{}

\begin{document}

\maketitle
\begin{abstract}
    Advanced space technology systems often face high fixed costs, can serve limited external demand, and are significantly driven by non-economic motivations. While increased entrepreneurial activity and national ambitions in space have encouraged planners at public space agencies to develop markets around such systems, the very factors that make the recent growth of the space economy so remarkable also challenge planners' efforts to develop and sustain markets for space-related goods and services. This paper proposes a diagram to visualize the number of competitors a market can sustain as a function of the industry's cost structure; the distribution of program support across direct purchases, direct investments, and shared infrastructure; and the magnitude of demand external to the program. Building on public goods theory, the diagram shows how marginal dollars invested in shared infrastructure can create non-rival benefits supporting more competitors than direct purchases or subsidies. The diagram is demonstrated with a stylized application inspired by NASA's Commercial LEO Destinations program. Under conditions consistent with public data, independent stations generate industry-wide losses of \$355 million annually, while shared core infrastructure enables industry-wide profits of \$154 million annually. The choice between shared infrastructure and direct purchases can depend on economic conditions that are outside the agency's control. Under favorable conditions---e.g., strong demand and low capital costs---direct purchases may suffice to sustain competition. Under challenging conditions---e.g., limited demand or tight capital markets---even well-designed shared infrastructure may prove insufficient to sustain competition.
\end{abstract}

\epigraph{Technology, under all circumstances, leads to planning; in its higher manifestations it may put the problems of planning beyond the reach of the industrial firm. Technological compulsions, and not ideology or political wile, will require the firm to seek the help and protection of the state.}{--- John Kenneth Galbraith, \textit{The New Industrial State}} % pp. 20

\section{Introduction}

Public space agencies are increasingly interested in developing sustainable competition to supply advanced space technologies to meet growing commercial and international interests. Several have adopted market development goals, in some cases seeking to expand their commercial procurement of space services while maintaining certain national industrial capabilities.\footnote{For example, public space agencies in the U.S., Japan, and Europe have all announced various types of market development goals \citep{nasa_nasa_2019, european_commission_eu_2025, esa_esa_2025, cabinet_office_government_of_japan_space_2024, ey_global_japan_2024}, while NASA's investment in the Space Launch System was influenced by industrial base considerations \citep{NASA_SLS_IndustrialBase_2011}.} As space ambitions rise globally while public budgets face greater strain, understanding and clearly communicating principles of pro-competitive public investment in advanced technology systems is an increasingly important task for economists and public space agency officials. Yet a basic economic question for market development---how should a program allocate its budget across purchases, transfers, and infrastructure to sustain competition to supply space-related goods and services---remains underexplored.

Space market development efforts face deep uncertainty. Measurements of the current space economy are contested, with independent analyses revealing that some industry projections systematically overstate growth by double-counting government revenues and input costs \citep{crane_measuring_2020}. Forecasting demand for markets that don't yet exist---like lunar surface power or commercial space stations---is more challenging still. Cost structures are often speculative, and beliefs about the relative efficiency of different contract types often lack rigorous econometric grounding. Factors like external demand (i.e., demand outside the agency's control) and interest rates can overwhelm program design choices, leaving agencies vulnerable to conditions beyond their control.

Despite this uncertainty, public space agencies must decide how to distribute program budgets across direct purchases, transfers, and shared infrastructure. This paper introduces a ``sustainable competition'' diagram showing how many firms can be sustained at different levels of program spending, external demand, and cost imposed by requirements, to help planners structure thinking about economic trade-offs and uncertainty in program design. The diagram shows boundaries between profitability regions as a function of the agency's choice variables, making it easy to assess program design considerations like whether particular expenditures can sustain the desired number of firms and how large a shock would be required to threaten the program's economic sustainability. Building on the standard theory of public goods \citep{samuelson_pure_1954}, the key insight is that marginal dollars invested in shared infrastructure can create non-rival benefits supporting more competitors than direct purchases or subsidies.

Space situational awareness illustrates this logic. The U.S. Space Force operates the Space Surveillance Network, a system of ground-based sensors and satellites tracking more than 49,000 objects in Earth orbit \citep{SpaceTrack_About}. The 18th Space Defense Squadron maintains this catalog and makes it freely available to U.S. and international satellite operators, academia, and other entities through space-track.org in order to reduce the risk of collisions between satellites. Public provision of basic tracking services as shared infrastructure lowers the cost of entry for commercial satellite operators, facilitating market development. Private firms can use the catalog to provide value-added analytics services without recreating the full network of ground-based sensors and satellites.

A stylized application based on NASA's Commercial LEO Destinations program formalizes this logic and illustrates use of the diagram. The application compares independent free-flyer versus shared core module architectures for sustaining at least two competing crewed space station operators. Under cost and demand conditions consistent with public data, independent stations generate industry-wide losses of \$355 million annually, while shared core infrastructure enables industry-wide profits of \$154 million annually.

Section 2 describes the general approach assessing economics of market development programs and constructing the diagram. Section 3 develops estimates and constructs the diagram for an application to crewed space stations in low-Earth orbit. Section 4 identifies key directions for future research on public investment and market development strategies for advanced technologies. Section 5 concludes.

\section{A diagram for assessing economics of market development programs}

Consider a public space agency with annual budget $B$ designing a program to develop sustainable markets for a class of advanced space technology systems. These systems---such as orbital debris tracking radars, lunar surface power sources, or orbital debris removal spacecraft---typically exhibit high fixed costs, long development timelines, and highly uncertain external demand, creating conditions where uncoordinated private investment may result in too few competitors or none at all. As \cite{weinzierl_space_2018} notes, complementarities between space technologies mean space businesses across the technology stack play stag hunts with each other. This analysis assumes the public space agency is a planner capable of using its program requirements and budget to structure technological complementarities between private actors to encourage desired outcomes. The $N$ firms that develop and use the technology systems of interest are referred to as ``the industry.'' All economic variables are expressed as equivalent annuity units to facilitate comparison across different payment timing structures, firms are assumed to be symmetric in costs and capabilities, and inflation is ignored.

The agency can use the program budget to purchase goods and services produced by the industry ($R^G_i$ from firm $i$), to invest in shared infrastructure for the industry ($G^S$), or to directly transfer funds to the industry ($G_i$ to firm $i$, $G^D = \sum_i G_i$).\footnote{The specific mechanisms, e.g., loans, grants, preferred shares, etc. are not relevant to the high-level analysis here and introduce additional complexity. Non-dilutive cash grants are a common transfer mechanism and analytically tractable.} Ignoring overhead and administrative costs, the program's budget constraint is:
\begin{equation}
\sum_i R^G_i + \sum_i G_i + G^S \leq B .\label{eqn:budget_constraint}
\end{equation}

Public investment into shared infrastructure provides benefits $Y^G$ to the industry according to the valuation function $f(G^S)$, which measures how much shared infrastructure reduces firms' costs relative to developing equivalent capabilities independently.\footnote{The specific functional form depends on the technology systems and business models involved. For modular systems, benefits may be approximately linear up to shareable essential element costs, then exhibit diminishing returns; for network systems requiring interoperability, benefits may show increasing returns to scale.} Here, firms are assumed to uniformly value shared infrastructure at cost: $Y^G = G^S$.

Firms in the industry earn total revenue $R_i$, with $R_i^G$ earned from the program and $R^M_i$ earned from sources external to the focal agency---individuals, households, firms, other public space agencies, and any other entities that may purchase the industry's output. This separates revenues the focal agency controls from those it cannot directly control; while the agency may coordinate with other entities, the agency does not control their spending. Under symmetry, each firm receives equal shares: $R^G_i = R^G / N$ and $R^M_i = R^M / N$. Similarly, direct transfers simplify to $G_i = G^D / N$.

Total industry revenue is $R = \sum_{i}(R_i^G + R_i^M)$. Firms incur gross total cost $X_i$ and net total cost $C_i = X_i - Y^G - G_i$; industry total cost is $C = \sum_i C_i$. The gross costs $X_i$ represent costs to develop, deploy, and operate the target system; under symmetry, $X_i = X/N$ where $X = \sum_i X_i$ is industry-level gross total cost.\footnote{Space technology program managers recognize a positive relationship between system performance and cost \citep{nasa_executive_cost_analysis_steering_group_nasa_2015}, allowing cost to proxy---with suitable caveats---for system performance required by the program.}

Assuming the agency exhausts its budget, individual firms' profits are:
\begin{align}
\Pi_i &= R_i - C_i \nonumber \\
% &= R_i^G + R_i^M - (X_i - Y^G - G_i) \nonumber \\
% &= \sigma_i^G(R^G) + \sigma_i^M(R^M) - (X_i - Y^G - G_i) \nonumber \\
% &= R^G/N + R^M/N - X/N + f(G^S) + G^D/N \nonumber \\
&= \frac{B - G^S + R^M - X}{N} + f(G^S) \label{eqn:profits}
\end{align}

A marginal dollar provides greater support to the industry if invested in shared infrastructure than if used for direct purchasing when:
\begin{align}
\forall i, ~\frac{\partial \Pi_i}{\partial G^S} &> \frac{\partial \Pi_i}{\partial R^G} \nonumber %, \nonumber \\
% \implies f_i'(G^S) &> \sigma_i'(R^G)
\end{align}

In the symmetric case, this reduces to:
\begin{align}
f'(G^S) &> \frac{1}{N} .\label{eqn:infrastructure_dominance_condition}
\end{align}

An identical condition holds for direct investment.

If the public space agency wants to maintain at least $N$ competitors, then it must ensure $R^G$, $G^D$, and $Y^G$---its support for the industry---are sufficient to ensure all $N$ firms earn non-negative profits, conditional on the anticipated level of external demand, $R^M$. Formally:
\begin{align}
\forall i, ~ \Pi_i(N) & \geq 0 \nonumber \\
\implies \frac{B - G^S + R^M - X}{N} + f(G^S) &\geq 0 \label{eqn:profitability_constraint}
\end{align}

This constraint captures the fundamental trade-off: supporting more competitors requires either reducing individual firm costs (through shared infrastructure) or increasing industry revenues (through direct purchases, transfers, or market growth). Programs violating this constraint will see firms exit, potentially ending in monopoly or no suppliers.\footnote{Questions regarding payment timing and conditioning can be investigated by appropriately modeling and discounting benefit flows when calculating $R^G$, $G^D$, and $Y^G$.}

Overlaying profitability regions across different industry sizes shows the full architectural possibility set. For fixed $N$, these regions are $\{(R^G, C): ~ \forall i, ~ \Pi_i(N) \geq 0 \}$---intersections of superlevel sets for firms' profitability constraints.\footnote{Similar diagrams are used in managing engineering costs and schedules in advanced technology programs, e.g., Joint Confidence Level plots \citep{nasa_executive_cost_analysis_steering_group_nasa_2015}.} Direct government purchases, $R^G$, represent agency support appearing on firms' income statements; industry total costs, $C = X - Y^G - G^D$, show the cost burden firms face after accounting for shared infrastructure benefits and direct transfers. Regions where profits are non-negative show combinations of costs and purchases that can sustain a given number of firms.

\begin{figure}[H]
\centering
\includegraphics[width=0.8\textwidth]{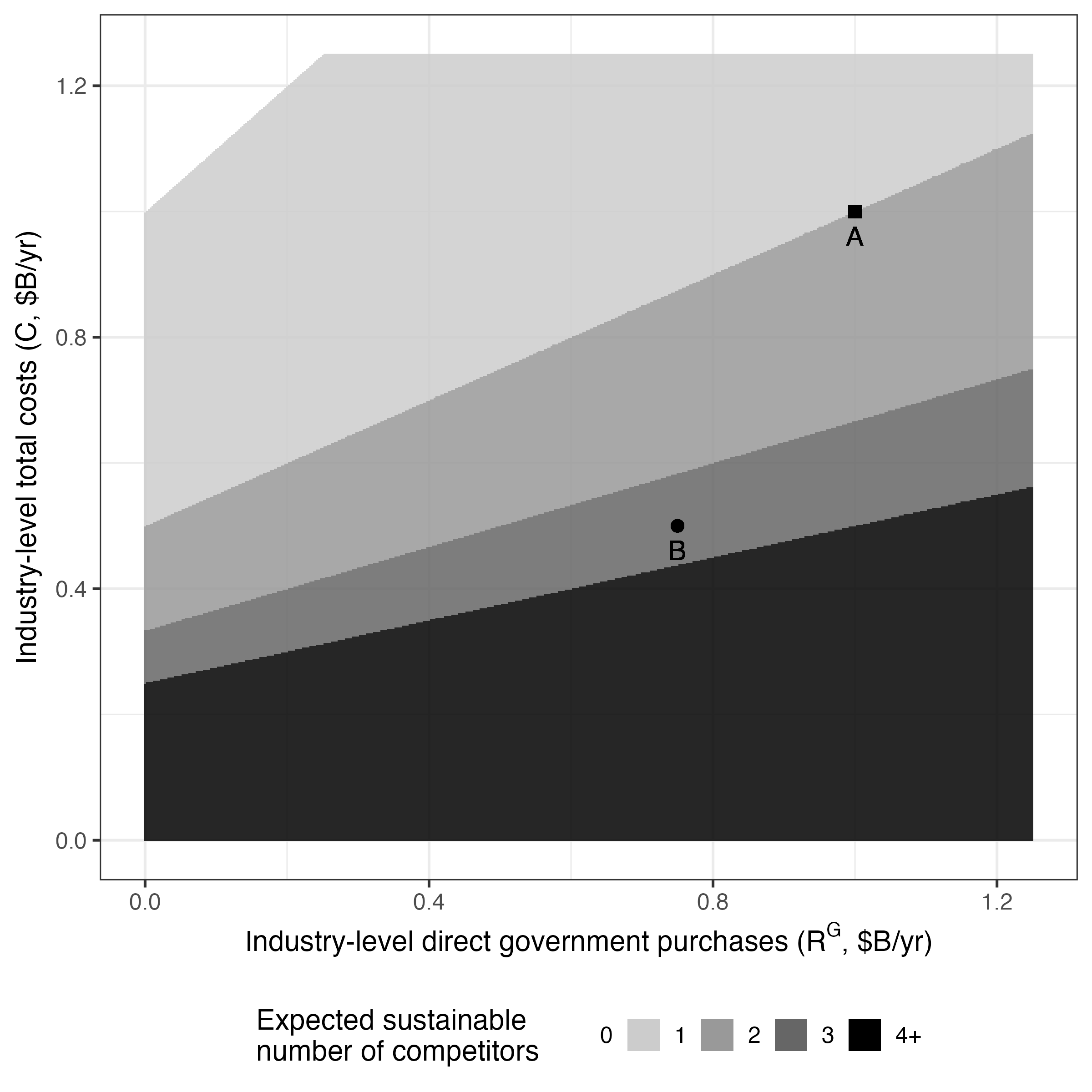}
\caption{Sustainable competition diagram showing the relationship between industry-level direct government purchases, industry-level total costs, and the potential number of profitable firms. Shaded regions indicate the expected sustainable number of competitors under different system cost and government-derived revenues.}
\label{fig:sustainable_competition}
\end{figure}

To illustrate, consider two architectures for sustaining at least two firms: architecture A transfers all funds as non-dilutive grants; architecture B divides its budget evenly between direct purchases and shared infrastructure. With external revenues of \$1B/year, gross costs of \$2B/year, program budget of \$1B/year, and firms valuing shared infrastructure at cost. Table \ref{table:econ-parameters-example} lists the economic parameters. Under architecture A the industry barely breaks even; under architecture B it earns positive profits. Figure \ref{fig:sustainable_competition} shows architecture A on the boundary between sustainable duopoly and monopoly regions, while architecture B lies within the sustainable triopoly region.

\begin{table}[htbp]
\centering
\caption{Economic parameters under alternative program architectures}
\label{table:econ-parameters-example}
\begin{tabular}{lll}
\toprule
 & \textbf{Economic architecture A} & \textbf{Economic architecture B} \\
\midrule
Desired number of competitors ($N$) & 2 firms & 2 firms \\
Program budget ($B$) & \$1 B/year & \$1 B/year \\
\hdashline
\quad \textit{Direct transfers} ($G_i$) & \textit{\$0.5 B/year/firm} & \textit{---} \\
\quad \textit{Direct purchases} ($R^G_i$) & \textit{---} & \textit{\$0.25 B/year/firm} \\
\quad \textit{Shared infrastructure spending} ($G^S$) & \textit{---} & \textit{\$0.5 B/year} \\
\hdashline
External revenue ($R^M_i$) & \$0.5 B/year/firm & \$0.5 B/year/firm \\
Shared infrastructure value ($Y^G$) & \textit{---} & \$0.5 B/year/firm \\
Gross total cost ($X_i$) & \$1 B/year/firm & \$1 B/year/firm \\
\midrule
% Total firm revenues ($R_i$) & \$1 B/year/firm & \$0.75 B/year/firm \\
% Total firm costs ($C_i$) & \$1 B/year/firm & \$0.5 B/year/firm \\
% Total firm profits ($\Pi_i$) & \$0 B/year/firm & \$0.25 B/year/firm \\
% \hdashline
Total industry revenues ($R$) & \$2 B/year & \$1.5 B/year \\
Total industry costs ($C$) & \$2 B/year & \$1 B/year \\
\textbf{Total industry profits} ($\Pi$) & \textbf{\$0 B/year} & \textbf{\$0.5 B/year} \\
\bottomrule
\end{tabular}
\end{table}

Both representations show that architecture A requires additional budget outlays following cost or demand shocks to prevent industry collapse to monopoly, while architecture B can weather greater uncertainty. The next section illustrates how to construct the diagram with an application.

\section{An application to crewed space stations}

Crewed space stations exhibit the high fixed costs, uncertain external demand, and opportunities for shared infrastructure that characterize many advanced space technology systems. Like the tracking catalog in the Space Situational Awareness case, crewed stations also have foundational infrastructure that can be shared across firms. Suppose a public space agency with a human spaceflight program intended to maintain continuous human presence in low-Earth orbit while developing sustainable markets for crewed space station services.\footnote{``Continuous presence'' has been a stated goal of U.S. space policy for some time \citep{nasa_nasas_2024}. While a classical economic planner may focus on goals derived from economic welfare maximization, non-economic goals are common in space policy, which is often meant to support a sense of national achievement. \cite{macdonald_long_2017} discusses this dimension of space economics in greater detail.} Should the agency support multiple independent ``free-flyer'' stations, each providing complete space station capabilities, or should it invest in shared infrastructure that multiple firms can use while developing their own specialized modules? Using cost estimates derived from analogous system elements and budget assumptions consistent with NASA planning, this application compares these two approaches to demonstrate the diagram's use in program design decisions.

A program supporting continuous human presence requires substantial investment in both station capabilities and crew operations, with private investors expecting returns over the asset's full lifetime. The analysis assumes the program covers transportation to and from the station as well as crew supplies and equipment. Since transportation and crew support costs remain roughly constant regardless of whether the program supports independent stations or shared infrastructure, the economically relevant comparison focuses on the ``capturable budget''---funds available for direct purchases, direct transfers, and shared infrastructure investment. Based on the 2026 NASA President's Budget Request for commercial space station services, this capturable budget amounts to approximately \$1B annually.\footnote{The 2026 NASA President's Budget Request contains roughly \$2B for crewed space station services in 2030, when the ISS is retired and private stations become operational \citep{nasa_fy_2025}. This includes approximately \$1.2B for transportation and resupply flights. The remaining \$800M represents the capturable portion available for station-related investments, rounded to \$1B for analytical convenience.} The analysis assumes external demand from research institutions, private companies, and other national space agencies and government entities provides an additional \$500M annually in total potential revenue for station operators.

The economic architecture choices regarding shared infrastructure stem from the separability of space station functions. Space stations provide two distinct types of capabilities: core functions that every crewed station requires, and habitat functions that differentiate market offerings. Core functions include life support systems capable of sustaining crew between resupply missions, power generation and distribution, command and data handling, and docking capabilities for crew and cargo vehicles. These systems must meet stringent safety and performance requirements regardless of the station's intended market---a facility serving government researchers needs the same fundamental life support capabilities as one targeting commercial customers.\footnote{The Environmental Control and Life Support System (ECLSS) on the ISS currently performs at about 42\% O$_2$ recovery from CO$_2$ air loop closure \citep{nasa_benefits_2022}. A station aimed at long-duration research missions may require higher recovery rates than one serving short-duration tourists, but both need reliable life support systems.} Habitat functions, by contrast, determine what activities crew can perform: research laboratories, manufacturing facilities, recreational amenities, or other specialized capabilities. A crewed station operator targeting pharmaceutical research will invest differently in habitat capabilities than one serving space tourism, but both require similar core infrastructure to keep humans alive and connected to Earth. The core functions represent foundational infrastructure---essential prerequisites without which crewed operations cannot occur---while habitat capabilities represent enabling infrastructure that expands opportunities for service differentiation. This functional separation creates a natural opportunity for shared infrastructure investment, where the public agency provides essential core capabilities that all operators need, while private firms differentiate their offerings through specialized habitat modules. This parallels the Space Situational Awareness case: the tracking catalog is foundational infrastructure, and value-added analytics are how Space Situational Awareness providers differentiate their offerings.

Suppose the program supports continuous presence through stations accommodating up to 8 crew members, with government flights carrying up to 2 agency astronauts per flight through two crew rotations annually, leaving at least 2 seats per flight available for other paying customers.\footnote{The analysis does not assume the government space agency fully fills each flight with its own astronauts, only that the agency pays for the flight. The public space agency may use the seats to barter with other space agencies, to sell back to the public at a discount, or find another use for them. Barter arrangements between space agencies using the ISS are not uncommon \citep{veldhuyzen_no_1999, selding_esa_2014, foust_nasa_2025}.} This flight cadence, supported by two dedicated resupply missions annually, ensures substantial station capacity remains available for non-government customers, creating the market opportunity that private investors must evaluate. Stations are assumed to have 15-year design lifetimes with construction beginning in 2025, and private investors are assumed to expect 5\% annual returns over the asset lifetime.\footnote{Depreciation and maintenance requirements will depend on the system architecture; the analysis assumes these are captured in the operating costs.} These costs and revenues are assumed to be fixed in base year terms for the duration of the program. In the independent free-flyer architecture, each firm develops complete station capabilities including both core and habitat functions. In the shared core module architecture, the government provides a common core infrastructure supporting life support, power, docking, and robotic arm capabilities, while private firms develop specialized habitat modules that connect to this shared foundation. The analysis assumes the shared core module can accommodate up to two habitat modules, with government demand for station services split equally between providers in scenarios with multiple competitors. These assumptions are consistent with current NASA use of the ISS and the capabilities of existing crew and cargo vehicles.

\begin{figure}[H]
\centering
\includegraphics[width=0.8\textwidth]{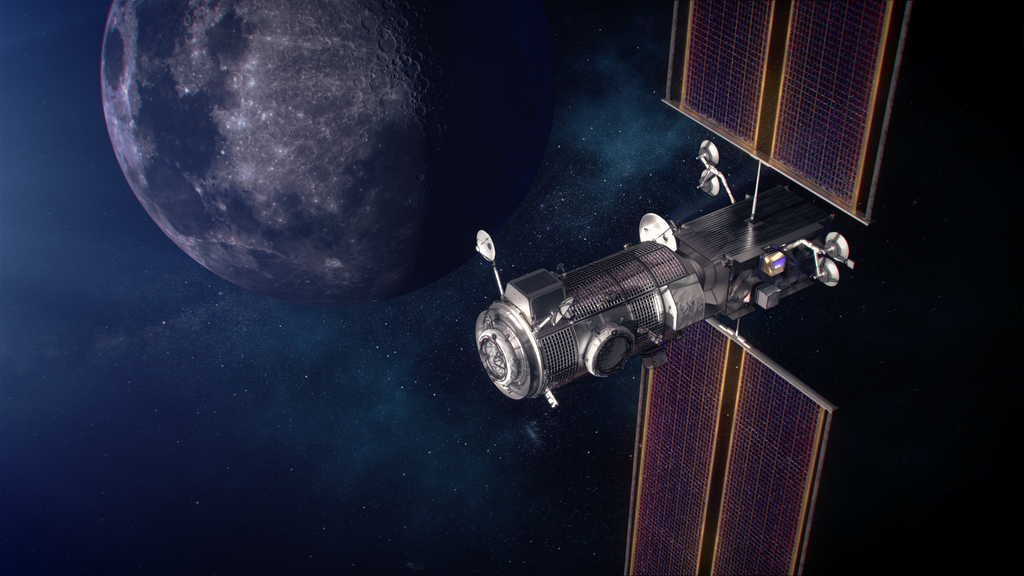}
\caption{Artist's concept of the Gateway's PPE and HALO elements in lunar orbit. The PPE is the rectangular structure to which solar panels are attached, while the HALO is the cylindrical structure with circular docking ports visible. Credit: NASA.}
\label{fig:halo-ppe}
\end{figure}

\begin{figure}[H]
\centering
\includegraphics[width=0.8\textwidth]{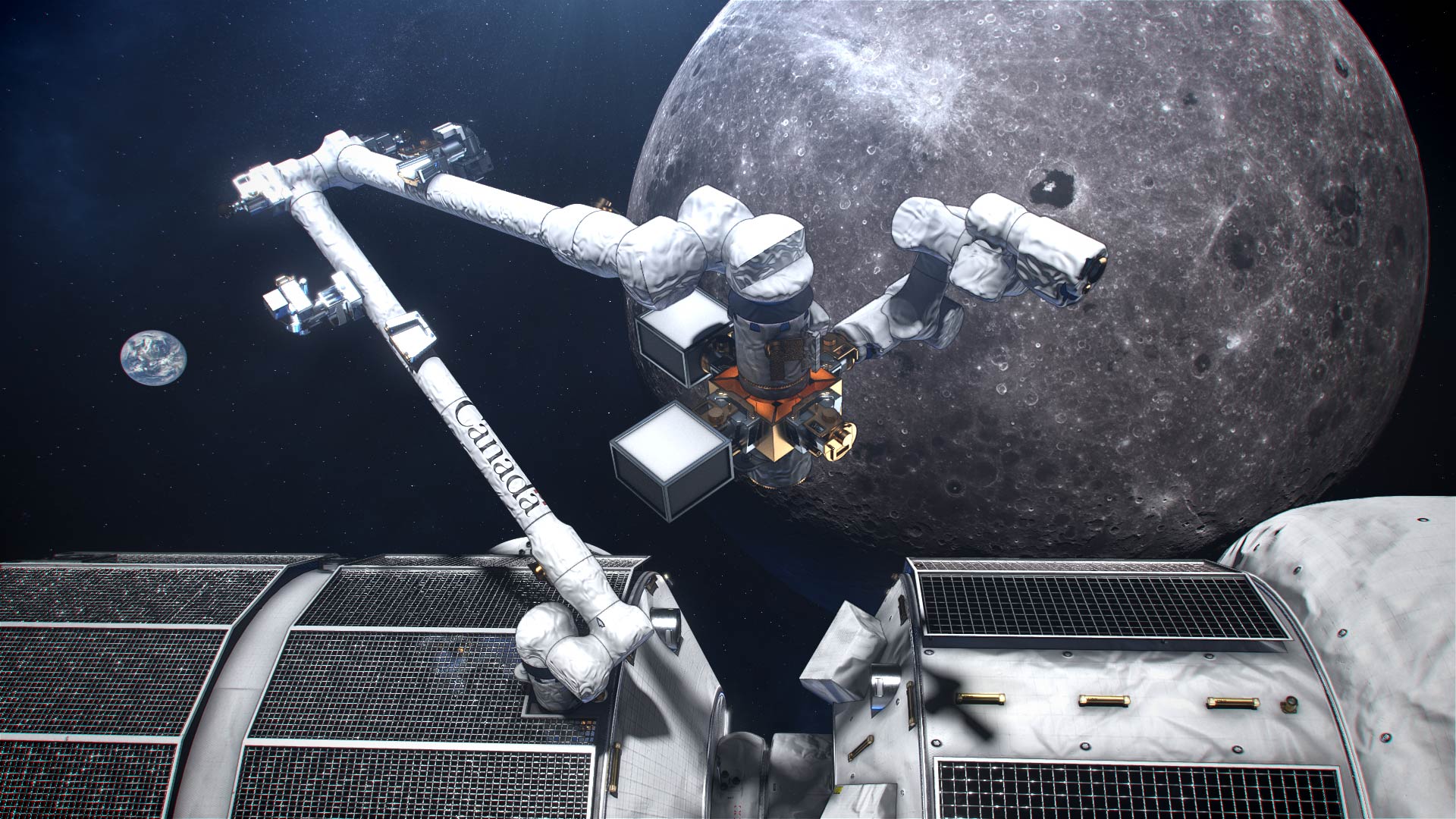}
\caption{Artist's concept of CSA's Canadarm3, an exterior robotic arm, on the exterior of Gateway. Credit: CSA.}
\label{fig:canadarm3}
\end{figure}

\begin{figure}[H]
\centering
\includegraphics[width=0.8\textwidth]{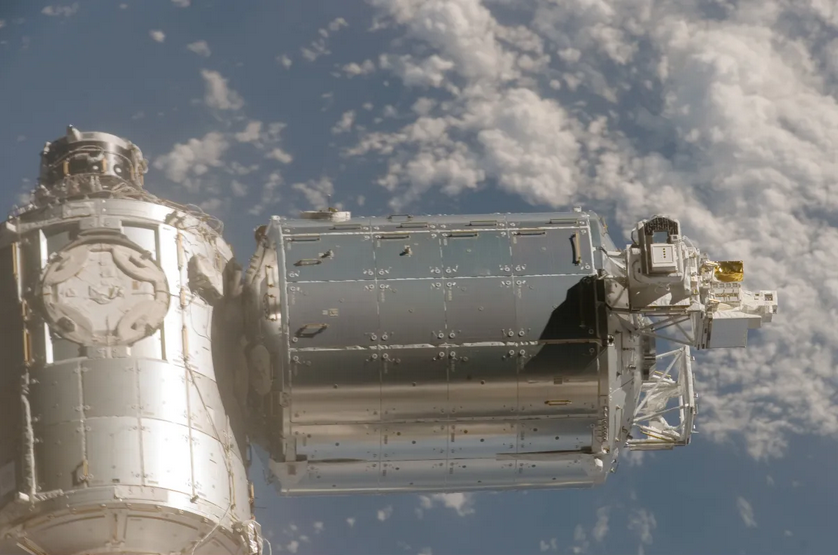}
\caption{ESA's Columbus laboratory module attached to the Harmony module on the ISS. Credit: ESA.}
\label{fig:columbus-exterior}
\end{figure}

To estimate the costs of these two architectures, cost data is derived from comparable space station components currently under development or currently operational. Cost estimates for core module elements come from NASA's Gateway program, which provides the closest available analogs for long-duration crewed space infrastructure: the Power and Propulsion Element (PPE) contract provides cost data for power and thermal systems, the Habitation and Logistics Outpost (HALO) contract provides data for pressurized volume and life support integration, and the Canadian Space Agency's (CSA) Canadarm3 contract provides data for robotic capabilities. Habitat module construction costs derive from the European Space Agency's (ESA) Columbus laboratory module, while operating costs are estimated using Columbus' share of total ISS habitable volume applied to current ISS operating expenses. All cost figures are adjusted to 2025 dollars using NASA's New Start Inflation Index, which accounts for space technology-specific cost escalation.\footnote{Several cost categories are excluded from this analysis. Insurance costs, which can be significant for crewed systems, are omitted due to indications that private insurance markets lack capacity to cover fully privately owned crewed space stations \citep{macdonald_commercial_2024}. Launch costs for initial system deployment are excluded, though including them would favor the shared core architecture. Training costs and supply-side structure details for crew operations are abstracted away, though transportation monopolies or inadequate training facilities could complicate the economics of human spaceflight.} Integration costs for the core module are assumed to be 15\% of total element costs. Figure \ref{fig:halo-ppe} shows an example of a core module, Figure \ref{fig:canadarm3} shows an example of a robotic arm, and Figure \ref{fig:columbus-exterior} shows an example of a habitat module.

Table \ref{tab:core-module-element-costs} shows a list of core elements comprising a core module (i.e., foundational infrastructure) as described and, by analogy to roughly similar systems, approximate costs. Table \ref{tab:habitat-module-costs} shows the costs for a habitat module (i.e., enabling infrastructure). All figures are expressed in millions of 2025 U.S. dollars and rounded to the nearest million.

\begin{table}[htbp]
\centering
\caption{Core module element cost breakdown. All values rounded to the nearest million.}
\label{tab:core-module-element-costs}
\begin{tabular}{p{4cm}p{2.5cm}p{7.5cm}}
\toprule
\textbf{Cost Component} & \textbf{Value} & \textbf{Source and Notes} \\
\midrule
PPE and TTC & \$453 M & Power distribution, thermal control, operational control for entire platform. Based on Gateway PPE contract \cite{nasa_nasa_2020}. Adjusted from 2020 U.S. dollars. \\
\hdashline
Structure and outfitting & \$1132 M & Node, module, shelter, docking, micrometeroid and orbital debris protection, internal pressurized volume outfitting. Based on Gateway HALO contract \citep{nasa_nasa_2021}. Adjusted from 2021 U.S. dollars. \\
\hdashline
Robotic arm & \$752 M & Station assembly, passive berthing, external operations. Based on Canadarm3 contract \citep{csa_canada_2024}. Adjusted from 2024 Canadian dollars. \\
\hdashline
Core module integration & \$351 M & Assembling and integrating core structure elements. Assumption that integration costs are 15\% the total cost of core elements. \\
\midrule
\textbf{Core module construction cost} & \textbf{\$2688 M} & \textbf{Sum of rows above.} \\
\hdashline
\quad \textit{Construction cost annuity} & \textit{\$259 M/year} & \textit{Annuitized over 15 years at 5\%.} \\
\hdashline
Operations cost & \$250 M/year & Based on the operations costs for a private space station elicited from experts in \cite{crane_market_2017}: ``One industry expert estimated that operations costs for a modular space station could be \$200 to \$300 million.''  \\
\midrule
\textbf{Core module total cost annuity} & \textbf{\$509 M/year} & \\
\bottomrule
\end{tabular}
\end{table}

\begin{table}[htbp]
\centering
\caption{Habitat module cost breakdown. All values rounded to the nearest million.}
\label{tab:habitat-module-costs}
\begin{tabular}{p{4cm}p{2.5cm}p{7.5cm}}
\toprule
\textbf{Cost Component} & \textbf{Value} & \textbf{Source and Notes} \\
\midrule
Construction cost & \$1934 M & Based on the Columbus laboratory module, including testing and integration \citep{dlr_european_2023}. Adjusted from 2008 Euros. Columbus only has research racks without necessary support systems for crew bunks. \\
\hdashline
\quad \textit{Construction cost annuity} & \textit{\$186 M/year} & \textit{Annuitized over 15 years at 5\%.} \\
\hdashline
Operations cost & \$232 M/year & Columbus has total volume of 75 m$^3$ \citep{esa_european_2025}. ISS has total habitable volume of 388 m$^3$ \citep{nasa_international_2025}. 2021 NASA OIG found ISS operating expenses steady at roughly \$1.2B/year over 2016-2020 \citep{nasa_nasas_2021}. FY2026 NASA President's Budget Request projects similar costs from 2025-2030 \citep{nasa_fy_2025}. Estimated as ISS total operating expenses scaled by Columbus' share of habitable volume: $(75/388) \times 1200$. \\
\midrule
\textbf{Habitat module total cost annuity} & \textbf{\$418 M/year} & \\
\bottomrule
\end{tabular}
\end{table}

\begin{figure}[H]
\centering
\includegraphics[width=0.8\textwidth]{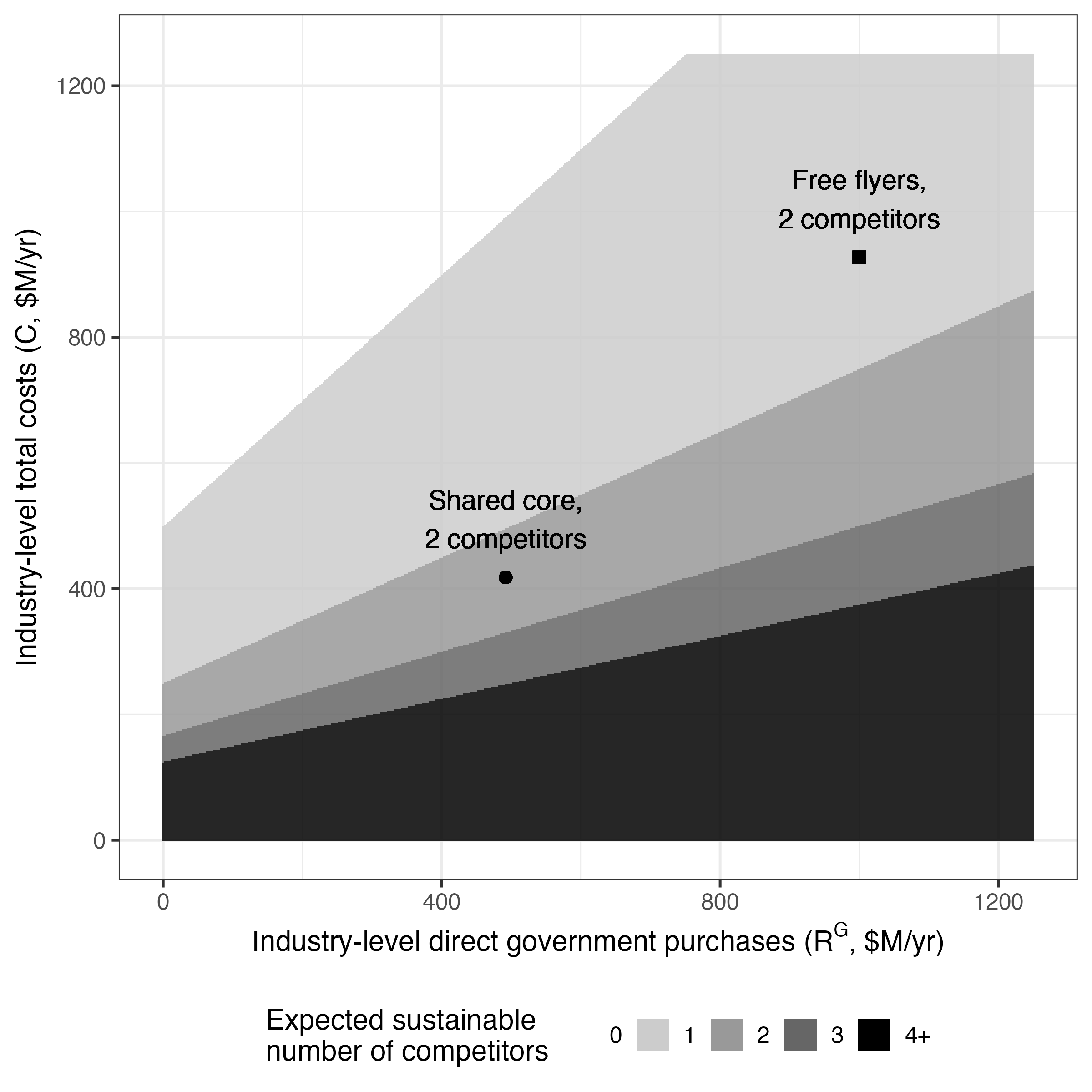}
\caption{Sustainable competition diagram showing the relationship between industry-level direct government purchases, industry-level total costs, and the potential number of profitable firms for the crewed space stations example. Shaded regions indicate the expected sustainable number of competitors under different system cost and government-derived revenues.}
\label{fig:sustainable-competition-stations}
\end{figure}

These cost estimates allow direct application of the Section 2 framework to assess how the choice between independent stations and shared infrastructure affects competitive viability. Both architectures face identical expected economic conditions: \$500M annually in external demand split between competitors, and \$1B in total capturable program budget. The architectures differ in how this budget gets allocated between direct purchases from firms versus investment in shared infrastructure. In the independent free-flyer architecture, the program budget flows entirely to direct purchases (\$500M per firm assuming two competitors). In the shared core architecture, the program invests in infrastructure---life support, power, docking, and robotic capabilities---that costs \$509M annually based on the component estimates in Table \ref{tab:core-module-element-costs}, leaving \$491M for direct purchases of habitat services (\$246M per firm). This reflects the functional equivalence of core capabilities across architectures. The shared core provides the same life support, power, and docking functions that firms would otherwise develop independently. This shared requirement is what makes core functions foundational and therefore a candidate for non-rival public provision. Firms are assumed to value the shared core functions identically whether provided by public infrastructure or developed independently. Under this assumption, firms in the shared core architecture benefit from \$509M in cost reduction while receiving \$246M in direct agency purchases.\footnote{The \$509M annual value represents what firms would need to spend to develop equivalent capabilities, not necessarily the agency's actual financing costs, which might differ due to existing assets, different financing terms, or economies of scale in procurement.} Table \ref{tab:econ-parameters-stations} presents the resulting economic parameters for both architectures, while Figure \ref{fig:sustainable-competition-stations} maps these onto the diagram to show how well each approach can sustain competition.

\begin{table}[htbp]
\centering
\caption{Economic parameters under both architectures. All values rounded to the nearest million.}
\label{tab:econ-parameters-stations}
\begin{tabular}{lll}
\toprule
 & \textbf{Independent free flyers} & \textbf{Shared core module} \\
 & \textbf{architecture} & \textbf{architecture} \\
\midrule
Desired number of competitors ($N$) & 2 firms & 2 firms \\
Program budget ($B$) & \$1000 M/year & \$1000 M/year \\
\hdashline
\quad \textit{Direct investment} ($G_i$) & \textit{---} & \textit{---} \\
\quad \textit{Direct purchases} ($R^G_i$) & \textit{\$500 M/year/firm} & \textit{\$246 M/year/firm} \\
\quad \textit{Shared infrastructure spending} ($G^S$) & \textit{---} & \textit{\$509 M/year} \\
\hdashline
External revenue ($R^M_i$) & \$250 M/year/firm & \$250 M/year/firm \\
Shared infrastructure value ($Y^G$) & --- & \$509 M/year \\
Gross total cost ($X_i$) & \$927 M/year/firm & \$927 M/year/firm \\
\midrule
Total industry revenues ($R$) & \$1500 M/year & \$991 M/year \\
Total industry costs ($C$) & \$1855 M/year & \$837 M/year \\
\textbf{Total industry profits} ($\Pi$) & \textbf{-\$355 M/year} & \textbf{\$154 M/year} \\
\bottomrule
\end{tabular}
\end{table}

Table \ref{tab:econ-parameters-stations} shows that under the assumed conditions, the independent free-flyer architecture generates negative industry profits of \$355M annually; two competing firms cannot both earn positive returns. Individual firms in this scenario face \$927M in annual costs while receiving only \$750M in combined program and external revenues (\$500M from program-driven purchases plus \$250M from external sources), resulting in \$177M annual losses per firm. The shared core architecture, by contrast, generates positive industry profits of \$154M annually, with each firm earning \$77M in annual profits after accounting for the \$509M cost reduction from sharing foundational infrastructure. Figure \ref{fig:sustainable-competition-stations} maps these results onto the sustainable competition diagram. The free-flyer architecture falls into the region where only monopoly provision is economically viable, while the shared core architecture can sustain two firms. That is, under the assumed conditions, a crewed free-flyer architecture is not compatible with competitive supply of habitat functions. Under the shared core approach firms earn positive profits, though not enough to incent entry of additional competitors. The choice of architecture thus determines whether competition to provide habitat functions is even possible, and the extent to which it may emerge.

\subsection{Exploring alternative assumptions}
\label{sec:application-scenarios}
The baseline analysis demonstrates how shared infrastructure can enable competition under a given set of conditions. The diagram can also be used to explore how different assumptions affect a program's economics. Two economic variables deserve particular attention because they lie largely outside space agency planners' control yet can significantly influence market development prospects: the scale of external demand for space station services, and the return expectations of private investors. External demand includes the broader commercial ecosystem for space-based research, manufacturing, and other applications---markets that may grow rapidly if early demonstrations prove successful, or remain limited if technical or economic barriers persist. Investor return expectations, meanwhile, fluctuate with broader macroeconomic conditions including interest rates, risk appetites, and alternative investment opportunities. By examining how the sustainable competition diagram changes under alternative values for these parameters, planners can assess the robustness of different economic architectures and identify conditions under which architectures might succeed.

\begin{figure}[H]
\centering
\includegraphics[width=0.8\textwidth]{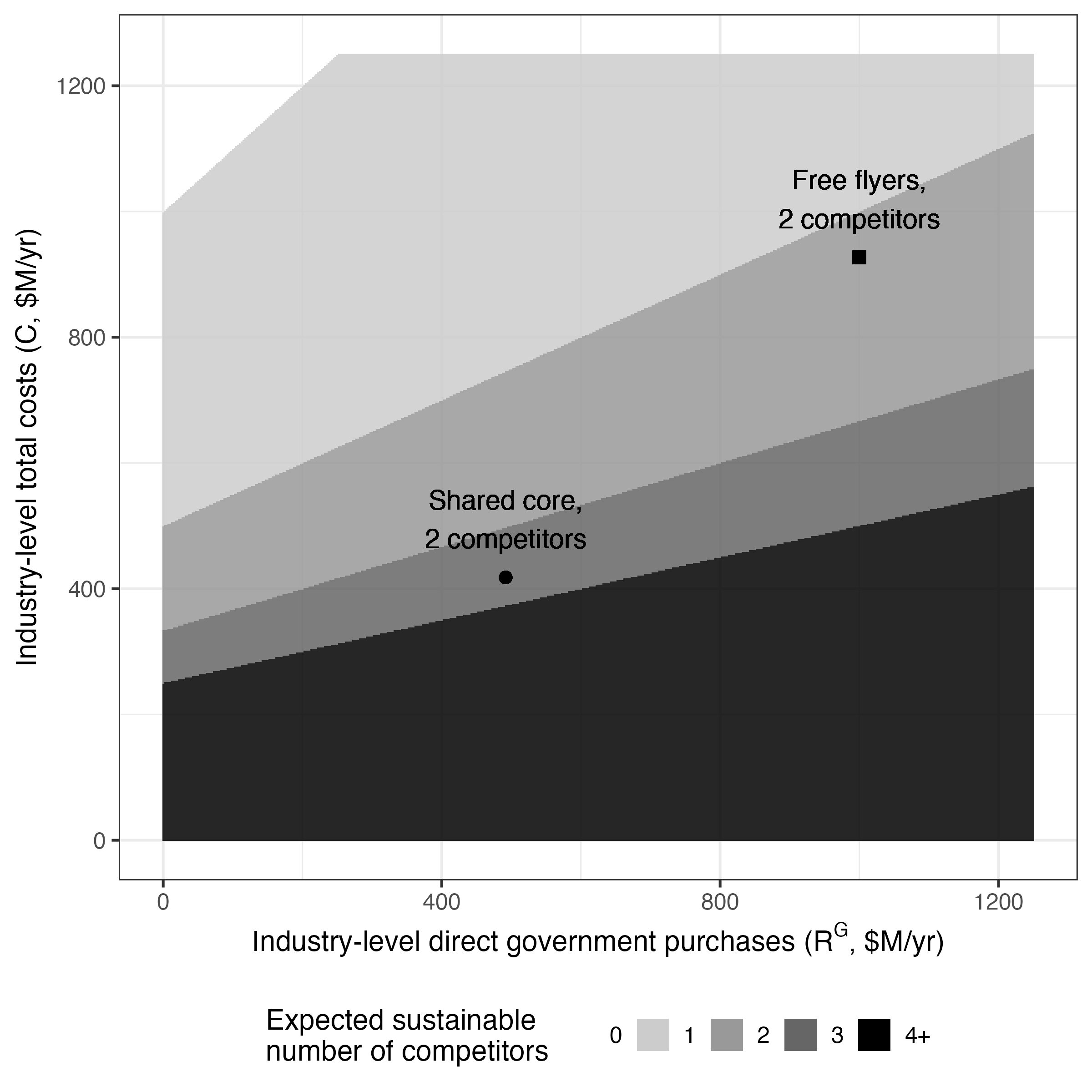}
\caption{Sustainable competition diagram showing the relationship between industry-level direct government purchases, industry-level total costs, and the potential number of profitable firms for the crewed space stations example with external demand of \$1000 M/year. Shaded regions indicate the expected sustainable number of competitors under different system cost and government-derived revenues.}
\label{fig:sustainable-competition-stations--high-demand}
\end{figure}

Figure \ref{fig:sustainable-competition-stations--high-demand} shows how doubling external demand from \$500M to \$1B annually affects both architectures. This scenario reflects an optimistic case where early space station demonstrations catalyze broader commercial adoption across research institutions, pharmaceutical companies, manufacturing firms, and other potential users. Under these conditions, both architectures become more viable and the choice of architectures becomes less significant. The independent free-flyer architecture, which generated \$355M in annual losses under baseline conditions, now produces positive industry profits and can sustain two competitors with margin for reinvestment or dividends. Each firm in the free-flyer scenario would earn \$500M annually from non-government customers plus \$500M from government purchases, covering the \$927M total costs with \$73M profit margin per firm. The shared core architecture performs even better, generating sufficient profits to sustain three competing firms rather than the two supported under baseline demand, with enough profits to provide returns to existing investors and potentially incent entry of additional competitors. However, new entrants would not be able to enter on the same terms as incumbents without more capacity on the shared core or a new core. This raises an important consideration in using shared infrastructure approaches: the design of the infrastructure may also impose limits on the number of firms that can be sustained under favorable economic conditions. Further analysis of this scenario can identify threshold levels of external demand such that direct purchases alone may sustain competition without requiring shared infrastructure. This outcome depends on successfully developing the commercial ecosystem---an uncertain prospect that may take years or decades to materialize, if ever.

\begin{figure}[H]
\centering
\includegraphics[width=0.8\textwidth]{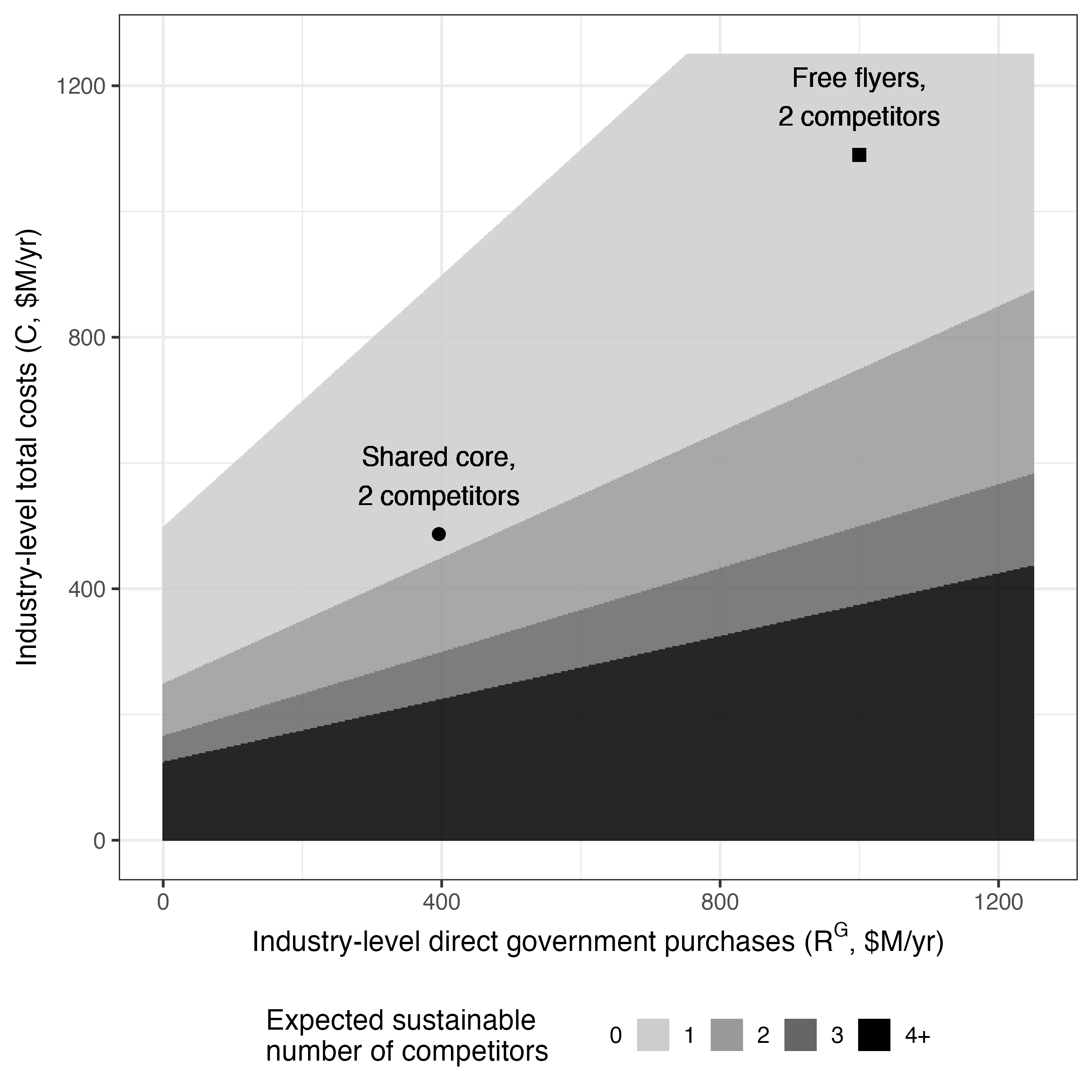}
\caption{Sustainable competition diagram showing the relationship between industry-level direct government purchases, industry-level total costs, and the potential number of profitable firms for the crewed space stations example with investor return expectations of 10\%/year. Shaded regions indicate the expected sustainable number of competitors under different system cost and government-derived revenues.}
\label{fig:sustainable-competition-stations--high-rates}
\end{figure}

Figure \ref{fig:sustainable-competition-stations--high-rates} shows how doubling investor return expectations from 5\% to 10\% annually affects both architectures. This scenario reflects tighter macroeconomic conditions where rising interest rates, increased risk premiums, or competition from alternative investments raise the hurdle rate for space technology projects. Under 10\% return requirements, neither architecture can sustain two competing firms. The independent free-flyer architecture, already unprofitable for two firms under baseline conditions, becomes even less viable with higher costs of capital. The shared core architecture, which supported two firms under baseline conditions, also falls into the monopoly-only region. If there were two firms in the shared core architecture, both would require higher annual profits to compensate investors for the increased return requirement. But revenues and infrastructure-driven cost reductions remain unchanged, and expected cashflows that allowed firms to raise capital under baseline conditions no longer appear as attractive. While space agencies are limited in their abilities to influence macroeconomic conditions, they can monitor them to proactively assess external headwinds or tailwinds and design programs accordingly. Agency-led efforts to aggregate and secure demand from other government entities may be particularly valuable when investors have many other lucrative-seeming opportunities to deploy large amounts of capital and non-government demand for services from habitat functions is weak.

These scenarios reveal several practical insights for public space agencies seeking to develop space markets. First, the choice between shared foundational infrastructure and direct purchases can depend on economic conditions that are outside the agency's control. Under favorable conditions---e.g., strong external demand and low capital costs---direct purchases may suffice to sustain competition without requiring shared infrastructure investments. Under challenging conditions---e.g., limited external demand or tight capital markets---even well-designed shared infrastructure may prove insufficient to maintain competitive supply. Partnerships with other space agencies to aggregate and channel demand are a potential mechanism to offset challenging external conditions. Second, given how external economic factors can dominate program design choices in determining market outcomes, timing market development efforts to coincide with favorable economic conditions may be as important as program design choices. Space agencies may benefit from integrating macroeconomic assessments into the design of programs that rely on private investors. Third, the diagram provides a systematic way to assess these trade-offs and communicate them to stakeholders, but depends on credible estimates of costs, external demand, and competitive dynamics that remain challenging to develop for novel space technology systems. The estimates may evolve with technical and economic conditions, requiring ongoing assessment and planning.

\section{Limitations and Directions for Future Research}
\label{sec:future-research}

The diagram developed here provides a systematic approach to assessing economic trade-offs and uncertainty facing planners designing programs with space market development objectives. Effective application requires expertise with space technology systems to construct relevant element costs, identify opportunities for efficient shared infrastructure, and assess whether candidate system architectures will meet public objectives. This analysis assumes firms are symmetric in costs and capabilities; allowing for asymmetry introduces additional modeling choices such as the order of entry or exit, quickly exploding the number of possible cases. Institutional knowledge can guide analysts toward particular asymmetric scenarios of interest.

The diagram may prove most useful as a way to formalize, consolidate, and communicate the various technical and economic assumptions that planners in public space agencies and in industry have adopted. The exercises conducted here may help economists see the kinds of insights public space agency planners may be able to use, while also helping public space agency planners see fundamental economic issues more clearly. Several research directions would enhance the practical applicability of economic analysis for advanced technology markets.

\textit{Cost differentials across contract types and governance structures.} Effective program design requires accurate assessment of cost differentials across economic architectures. This analysis assumed both economic architectures under consideration had identical costs. However, some in the space sector argue that firm fixed-price (FFP) contracts or private development result in lower costs than cost-plus fixed fee (CPFF) contracts \citep{berger_weirdly_2023}. Empirical evidence complicates these arguments. \cite{kim_counting_2025} found NASA has paid increasing real costs for launch services despite using FFP contracts with providers competing in markets where NASA is one among many customers. \cite{carril_impact_2020} found the U.S. Department of Defense's transition from FFP to CPFF contracts during a period of industry consolidation did not statistically significantly increase costs. \cite{kim_and_2025} found industry-led projects show statistically significant cost advantages over government-led projects only for low-risk systems like small satellites. FFP contracts may also create ``loser's curse'' effects: if contractors can exit when costs exceed expectations, FFP mechanisms transfer upside to contractors while leaving agencies exposed to downside risk.\footnote{For example, in 2022 NASA awarded Collins Aerospace a FFP contract under the xEVAS program to develop and produce space suits for ISS missions. Collins withdrew from the contract in 2024, reportedly due to cost and schedule overruns \citep{foust_collins_2022, foust_collins_2024}. Some observers have speculated that Boeing's Starliner project may face a similar fate \citep{terlep_boeing_2025}.} Credible econometric analysis of contract type and governance effects on costs could improve acquisition strategies and economic architecture assessments.

\textit{Investor perceptions of different revenue sources.} This analysis assumed that investors view all revenues, whether derived from a government or non-government source, identically. Is this a reasonable assumption? How do investor perceptions of space-derived cashflows actually vary with market and government budget volatility? Do investors systematically value government contracts differently from commercial revenues? Analyses like event studies of public space company stock prices around contract announcements, surveys of private space investors about their investment evaluation criteria, and choice experiments to elicit investor preferences about uncertain revenue streams could provide quantitative estimates of any systematic differences in how investors perceive different revenue sources. Such evidence would allow more accurate modeling of capital costs and return requirements across different economic architectures.

\textit{External demand forecasting.} Assessing economic trade-offs in program design requires estimates of external demand. For nascent markets like commercial space stations, external demand is deeply uncertain across multiple dimensions, e.g., how much demand will materialize, what services customers will demand, and when they will demand them. In the near term, external demand for advanced space systems may consist primarily of other government entities whose budgets and priorities the focal agency cannot control. Over longer timescales, the scale and composition of private commercial demand remains uncertain. Different customer types demand different service profiles: government agencies typically prioritize long-duration missions that signal capability and prestige, while commercial customers may prefer shorter-duration access for specific applications. Improved methods for forecasting demand under deep uncertainty---including both the scale of demand from various sources and the nature of services demanded---would help planners design better market development programs.

\textit{Applications to other domains.} This kind of economic analysis can be applied beyond crewed space stations. \cite{triezenberg_assessing_2020, triezenberg_assessing_2024} provide detailed analyses of similar goals and trade-offs in the U.S. National Security Space launch market, showing how acquisition decisions affect the number of sustainable competitors. Orbital debris risk reduction programs offer another example. While grab-and-remove active debris removal (ADR) systems can be cost-ineffective at reducing risk \citep{colvin_cost_2023, locke_cost_2024}, governments may want to support such technologies for non-economic reasons---these capabilities involve intermediate technologies with potential military applications. \cite{rao_cost-benefit_2025} assessed the economic efficiency of portfolios of space sustainability investments, identifying efficient combinations that include grab-and-remove ADR alongside ground-based lasers for small debris removal, new tracking systems, and improved shielding. Public ownership may be necessary for ground-based laser systems and some tracking systems that provide non-rival benefits to all satellite operators. Direct investments and purchases could support private firms developing ADR services or shielding. The economic architecture issues mirror those in the space station case: budget allocation, external demand assessment, risk distribution, and accounting for relevant non-economic factors.

\section{Conclusion}

This paper addresses a fundamental challenge facing public space agencies: how to allocate limited budgets between direct purchases, direct transfers, and shared infrastructure investments to develop and sustain markets for advanced technology systems. The core economic insight---that shared infrastructure creates non-rival benefits that can sustain more competitors than direct support mechanisms---follows from standard principles of public economics \citep{samuelson_pure_1954}. The crewed space station application demonstrates this: under plausible conditions, public investment in a shared core module enables two competing firms to earn substantial profits, while allocating that budget entirely toward direct purchases results in industry-wide losses. The sustainable competition diagram makes these trade-offs transparent to decision-makers.

The diagram can be applied broadly to advanced technology domains with high fixed costs, uncertain external demand, non-economic motivations, and separable system functions enabling shared infrastructure. Public science and technology agencies in such domains may face similar tensions between procurement objectives and industrial base or market development goals, while private investors must evaluate opportunities where government decisions can matter more than market fundamentals. On the other hand, the scenario analyses reveal that economic conditions can dominate program design choices in determining market outcomes. This highlights both the potential and the limits of public infrastructure investments for market development.

This logic can be extended to international cooperation. Infrastructure shared across partners reduces duplicated fixed costs, implying efficiency gains from cooperation. The Lunar Gateway exemplifies this: NASA provides power and propulsion, ESA contributes habitation and refueling modules, JAXA provides life support systems, Canada contributes a robotic arm, and the UAE provides the crew airlock, distributing development costs while enabling all partners to access common capabilities \citep{NASA_Gateway_2024}. Yet national interest in space systems is often driven by non-economic motivations, and geopolitical considerations can dominate efficiency calculations. China developed Tiangong independently after the Wolf Amendment effectively excluded Chinese participation in the ISS. Europe developed an independent satellite navigation system, Galileo, despite GPS being freely available. Fragmentation of infrastructure investment can also fragment external demand, since partners in shared infrastructure can become natural customers for each other's commercial providers. While economic analysis may not be able to resolve whether efficiency gains from cooperation outweigh the non-economic benefits planners perceive from strategic autonomy, it can clarify the economic stakes of those choices.

Market development plans require economic analysis grounded in the decisions planners actually face. Crewed space stations are a convenient example but not the only one. As global investment in advanced technologies accelerates, frameworks that help planners assess economic trade-offs and uncertainty will become more valuable to public and private technology investors.

\newpage
\section*{Declarations}
No funding was received for conducting this study.

\bibliography{references}

\end{document}